\documentclass[10pt, twocolumn]{article}

\usepackage[vlined, ruled, linesnumbered, noend]{algorithm2e}
\usepackage{lhelp}
\usepackage{hyperref}
\usepackage{graphicx}
\usepackage{amsmath}
\usepackage{amsfonts,latexsym,amssymb}
\usepackage{authblk}
\usepackage{lipsum}
\usepackage{color}
\usepackage{algpseudocode}

\newtheorem{Theorem}{Theorem}

\def\N {\ensuremath{\mathbb{N}}}

\def\Z {\ensuremath{\mathbb{Z}}}

\newcommand{\correctionhighlighted}[1]{{{\textcolor{red}{#1}}}}
\newcommand{\correctionnothighlighted}[1]{{{{#1}}}}

\newif\ifhighlighted
\highlightedtrue

\def\variable{2}

\newcommand{\correction}[2]{{{ \ifnum#1>\variable 
                                   \correctionhighlighted{#2}
                                \else
                                    \correctionnothighlighted{#2}
                                \fi
                              }}}

\newif\ifcomment
\commentfalse

\newif\iflongversion
\longversionfalse

\newif\ifthesis
\thesisfalse

\newif\ifnewversion
\newversiontrue

\newif\ifACM
\ACMfalse

\newif\ifrevision
\revisiontrue

\newif\ifcorr
\corrtrue

\begin{document}

\title{Parallel Integer Polynomial Multiplication}

\author{Changbo Chen}
\affil{changbo.chen@hotmail.com, CIGIT, Chinese Academy of Sciences}

\author{Svyatoslav Covanov}
\affil{svyatoslav.covanov@gmail.com, University of Lorraine, France}

\author{Farnam Mansouri}
\affil{mansouri.farnam@gmail.com, Microsoft Canada, Inc.}

\author{Marc Moreno Maza}
\affil{moreno@csd.uwo.ca, University of Western Ontario, Canada}

\author{Ning Xie}
\affil{nxie6@csd.uwo.ca, University of Western Ontario, Canada}

\author{Yuzhen Xie}
\affil{yuzhenxie@yahoo.ca, Critical Outcome Technologies, Inc.}

\maketitle

\begin{abstract}
\ifnewversion
We propose a new algorithm for multiplying dense
polynomials with integer coefficients in a parallel fashion, targeting
multi-core processor architectures.
Complexity estimates and
experimental comparisons demonstrate the advantages of this new
approach. 
\fi
\end{abstract}

{\em \textbf{Keywords.}
Polynomial algebra; symbolic computation;
parallel processing; cache complexity; multi-core architectures;}

\section{Introduction}
\label{sec:introduction}

Polynomial multiplication and matrix multiplication are at the
core of many algorithms in symbolic computation.
Expressing, in terms of multiplication time, 
the algebraic complexity of an operation,
like univariate polynomial division or the computation of a 
characteristic polynomial, is a standard
practice, see for instance the landmark book~\cite{Gathen:2003:MCA:945759}.
At the software level, 
the motto ``reducing everything to multiplication'' 
is also common, see for instance the
computer algebra systems
Magma~\cite{MR1484478}, 
NTL~\cite{shoup2001ntl} and
FLINT~\cite{Hart11flintfast}. 

With the advent of hardware accelerator technologies, such as
multicore processors and Graphics Processing Units (GPUs),
this reduction to multiplication is of course still desirable, but becomes
more complex, since both algebraic complexity and parallelism
need to be considered when selecting and implementing a multiplication algorithm.
In fact, other performance factors, such as cache usage or CPU pipeline optimization,
should be taken into account on modern computers, even on single-core processors.
These observations guide the developers of projects like 
SPIRAL~\cite{Pueschel:05}
or FFTW~\cite{Frigo05thedesign}.

This paper is dedicated to the multiplication of dense univariate
polynomials with integer coefficients targeting multicore processors.
We note that the parallelization of sparse (both univariate and multivariate) 
polynomial multiplication on those architectures has already been studied
by Gas\-tineau \& Laskar in~\cite{DBLP:conf/casc/GastineauL13}, and by
Monagan \& Pearce in~\cite{DBLP:conf/issac/MonaganP09}.
From now on, and throughout this paper, we focus on dense polynomials.
The case of modular coefficients was handled 
in~\cite{DBLP:conf/hpcs/MazaX09, DBLP:journals/ijfcs/MazaX11}
by techniques based on multi-dimensional FFTs.
Considering now integer coefficients, one can reduce
to the univariate situation through Kronecker's 
substitution, see the implementation techniques
proposed by Harvey in~\cite{Harvey:2009:FPM:1568787.1568920}.

A first natural parallel solution for multiplying
univariate integer polynomials is to consider 
divide-and-conquer algorithms where arithmetic counts
are saved thanks to distributivity of multiplication over addition.
{Well-known} instances of this solution are the multiplication algorithms
of Toom \& Cook, among which Karatsuba's method is a special case.
As we shall see with the experimental results of 
Section~\ref{sec:experimentation}, this is a 
practical solution. However, the parallelism is limited
by the number of ways in the recursion.
Moreover, augmenting 
the number of ways increases data conversion costs
and makes implementation 
quite complicated, see the work by Bodrato and Zanoni
for the case of integer multiplication~\cite{DBLP:conf/issac/BodratoZ07}.
As in their work, our implementation includes the  4-way and 8-way cases.
In addition, the algebraic complexity of an $N$-way Toom-Cook 
algorithm is not in the 
desirable complexity class of algorithms based on FFT techniques.

Turning our attention to this latter class, we first considered
combining Kronecker's substitution (so as to reduce
multiplication in ${\Z}[x]$ to multiplication in ${\Z}$)
and the algorithm of Sch{\"o}nhage \& Strassen~\cite{DBLP:journals/computing/SchonhageS71}.
The GMP library~\cite{granlund2015gnu} provides indeed
a highly optimized implementation of this latter algorithm~\cite{DBLP:conf/issac/GaudryKZ07}.
Despite of our efforts, we could not obtain much parallelism
from the Kronecker substitution part of this approach.
It became clear at this point that, in order to go beyond
the performance (in terms of arithmetic count and parallelism) of
our parallel $8$-way Toom-Cook code, our multiplication code had 
to rely on a parallel implementation of FFTs.

Based on the work of our colleagues from the SPIRAL and FFTW
projects, and based on our experience on the
subject of FFTs~\cite{DBLP:conf/hpcs/MazaX09,DBLP:journals/ijfcs/MazaX11,DBLP:conf/cap/MengVJMFX10},
we know that an efficient way to parallelize FFTs
on multicore architectures is the so-called {\em row-column} 
algorithm~\cite{VanLoan:1992:CFF:130635}.
which implies to view 1-D FFTs as multi-dimensional FFTs
and thus {differs from  the approach of Sch{\"o}nhage \& Strassen}.

Reducing polynomial multiplication in ${\Z}[y]$ 
to multi-dimen\-sional FFTs over a finite field, say ${\Z}/p{\Z}$, implies
transforming integers to polynomials over ${\Z}/p{\Z}$. 
Chmielowiec in~\cite{Andrzej13} experimented a similar method combined with 
the Chinese Remaindering Algorithm (CRA).
However, as we shall see, using a {\em big prime approach}
instead of a {\em small primes approach} opens
the door for using ``faster FFTs'' and reducing
algebraic complexity w.r.t to a CRA-based scheme.

As a result of all these considerations, we obtained
the algorithm that we propose in this paper.
We stress the fact that our purpose was not
to design an algorithm asymptotically optimal 
by some complexity measures.
Our purpose is to provide a parallel solution
for dense integer polynomial multiplication
on multicore architectures.
In terms of algebraic complexity, our algorithm
is asymptotically faster than that of 
{Sch{\"o}nhage \& Strassen}~\cite{DBLP:journals/computing/SchonhageS71}
while being asymptotically slower than that of 
F\"urer~\cite{DBLP:journals/siamcomp/Furer09}.
Our code is part of the {\em Basic Polynomial Algebra Subprograms}
and is publicly available at {\small \url{http://www.bpaslib.org/}}.

Let $a(y), b(y) \in {\Z}[y]$ and $d$ be a positive integer 
such that $d-1$ is the maximum degree of $a$ and $b$,
that is, $d = {\max}({\deg}(a), {\deg}(b)) + 1$.
We aim at 
computing the product $c(y) := a(y)$ $b(y)$.
We propose an algorithm whose principle is sketched below.
A precise statement of this algorithm is given in 
Section~\ref{sec:cmmxxmultiplication},
\iflongversion
while complexity results and implementation techniques 
appear in
Sections~\ref{sec:cmmxxmultiplication-complexity} and
\ref{sec:cmmxxmultiplication-implementation}.
\else
while complexity results  and experimental results
appear in Sections~\ref{sec:cmmxxmultiplication-complexity}
and \ref{sec:experimentation}.
\fi

\begin{enumerateshort}
\item Convert $a(y), b(y)$  to bivariate polynomials
       $A(x,y), B(x,$ $y)$ over ${\Z}$ (by converting
       the integer coefficients of  $a(y), b(y)$ to univariate polynomials 
       of ${\Z}[x]$, where $x$ is a new variable) such that 
        $a(y) = A({\beta}, y)$ and $b(y) = B({\beta}, y)$ 
       hold for some ${\beta} \in {\Z}$ (and, of course, such that
       we have 
       ${\deg}(A,y) = {\deg}(a)$ and ${\deg}(B,y) = {\deg}(b)$).
\item Let $m > 4 H$ be an integer, 
      where $H$ is the maximum absolute value of the coefficients
      of the integer polynomial $C(x,y) := A(x,y) B(x,y)$.
      The positive integers $m$, $K$ and  the polynomials $A(x,y), B(x,y)$
      are built such that the polynomials
      $C^{+}(x,y) := A(x,y) B(x,y) \mod{\langle x^K + 1 \rangle}$
      and $C^{-}(x,y) := A(x,y) B(x,y) \mod{\langle x^K - 1 \rangle}$
      are computed 
      over ${\Z}/m{\Z}$ via FFT techniques, while the following equation holds over {\Z}: 
\begin{equation}
\label{eq:recoveringC}
C(x,y) \, = \, - \frac{C^{+}(x,y)}{2} (x^K -1) \, + \, \frac{C^{-}(x,y)}{2} (x^K +1).
\end{equation}
\item Finally, one recovers the product $c(y)$ by evaluating the above equation
at $x = {\beta}$.
\end{enumerateshort}
Of course, the polynomials $A(x,y), B(x,y)$ are also constructed such that
their total bit size is proportional to that of $a(y), b(y)$, respectively.
In our software experimentation, this proportionality factor ranges between 2 and 4.
Moreover, the number ${\beta}$ is a power of $2$
such that evaluating the polynomials $C^{+}(x,y)$ and $C^{-}(x,y)$
(whose coefficients are assumed to be in binary representation)
at $x = {\beta}$ amounts only to addition and shift operations.

Further, for our software implementation on 64-bit computer architectures, the
number $m$ can be chosen to be either one machine word size prime $p$,
a product $p_1 p_2$ of two such primes
or a product $p_1 p_2 p_3$ of three such primes.
Therefore, in practice, the main arithmetic cost of the whole
procedure is that of two, four or six convolutions, those latter
being required for computing  $C^{+}(x,y)$ and $C^{-}(x,y)$.
All the other arithmetic operations (for constructing
$A(x,y), B(x,y)$ or evaluating the polynomials $C^{+}(x,y)$ and $C^{-}(x,y)$)
are performed in a single or double fixed precision
at  a cost which is proportional to that of reading/writing
the byte vectors representing $A(x,y)$, $B(x,y)$, $C^{+}(x,y)$ and $C^{-}(x,y)$.

Theorem~\ref{thrm:complexity} below
gives  estimates 
for the work, the span and the cache complexity 
of the above algorithm.
Recall that our goal is not to obtain an algorithm which
is asymptotically optimal for one of these complexity
measures. 
Instead, our algorithm is designed to be practically
faster, on multi-core architectures, 
than the other algorithms that are usually
implemented for the same purpose of multiplying
dense (univariate) polynomials with integer coefficients.

\ifnewversion
\smallskip
\begin{Theorem}
\label{thrm:complexity}
Let $N$ be a positive integer such that every coefficient 
of $a$ or $b$ can be written with at most $N$ bits. 
Let $K, M$ be two positive integers greater than 1 and
such that $N = K M$.
Assume that $K$ and $M$ are  
functions of $d$ satisfying the following asymptotic
relations $K \in {\Theta}(d)$ and $M \in {\Theta} (\log{d})$.
Then, the above algorithm for
multiplying $a(y)$ and $b(y)$ has
a work of $O(d K M {\log}(d K) \log\log(\log(d)))$
     word operations,
a span of $O({\log}_2(d) K M)$ 
      word operations and incurs
$O(1 + (d M K/L) (1 + {\log}_{Z}(d M K)))$
cache misses.
\end{Theorem}
\smallskip

A detailed proof of this result is elaborated in
Section~\ref{sec:cmmxxmultiplication-complexity}.
The assumptions $K \in {\Theta}(d)$ and $M \in {\Theta} (\log{d})$
are not strong. It is, indeed, possible to reduce
to this situation by means of the {\em balancing 
techniques}\footnote{\correction{1}{Suppose that $N \in \Theta(d \log(d))$
does not hold, thus, preventing us from choosing
$K, M$ such that $K \in {\Theta}(d)$ and $M \in {\Theta} (\log{d})$
both hold. Then, applying Kronecker's substitution (forward and backward)
we replace $(a, b)$ by a new pair of polynomials
of ${\Z}[y]$ for which $N \in \Theta(d \log(d))$ does
hold.}} presented in~\cite{DBLP:journals/ijfcs/MazaX11}.
Applying those techniques would only increase the
work by a constant multiplicative factor, typically
between 2 and 4.

It follows from this result that our proposed 
algorithm is asymptotically faster
than combining 
Kronecker's substitution and Sch{\"o}nhage \& Strassen.
Indeed, with the notations of Theorem~\ref{thrm:complexity},
this latter approach runs in $O(d N \log(d N) \log \log(d N))$
machine word operations.

While it is possible to obtain a poly-log time for the span
this would correspond to an algorithm with high parallelism overheads.
Hence, we prefer to state a bound corresponding to our implementation.
By using multi-dimensional FFTs, we obtain a parallel algorithm
which is practically efficient, as illustrated
by the experimental results of Section~\ref{sec:experimentation}.
In particular, the algorithm scales well.
In contrast, parallelizing a $k$-way Toom Cook
algorithms (by executing concurrently the point-wise
multiplication, see \cite{farnamThesis}) yields
only a ratio work to span in the order of $k$, that is,
a very limited scalability.
Finally, our cache complexity estimate is sharp.
Indeed, we control finely all intermediate steps with this respect,
see Section~\ref{sec:cmmxxmultiplication-complexity}.

To illustrate the benefits of a parallelized dense univariate polynomial
multiplication, we integrated our code into the univariate real root
isolation code presented in~\cite{CMX11} together with a parallel
version of Algorithm (E) from \cite{DBLP:conf/issac/GathenG97}
for Taylor shifts. The results reported in~\cite{chen2014basic}
show that this integration has substantially improved the performance of our
real root isolation code.

\else

\fi

\section{Multiplying integer polynomials via two convolutions}
\label{sec:cmmxxmultiplication}

\iflongversion
\paragraph{Notations.}
\fi

We write 
\iflongversion
\begin{equation}
a(y) = \sum_{i=0}^{d-1} \, a_i y^i,  \ \ 
b(y) = \sum_{i=0}^{d-1} \, b_i y^i  \ \ {\rm and} \ \ 
c(y) = \sum_{i=0}^{2d-2} \, c_i y^i,
\end{equation}
\else
$a(y) = \sum_{i=0}^{d-1} \, a_i y^i$,
$b(y) = \sum_{i=0}^{d-1} \, b_i y^i$ and 
$c(y) = \sum_{i=0}^{2d-2} \, c_i y^i$, 
\fi
where $a_i, b_i, c_i$ are integers and $c(y) = a(y) b(y)$.
Let $N$ be a non-negative integer such that each coefficient ${\alpha}$
of $a$ or $b$ satisfies
\begin{equation}
\label{eq:definingN}
-2^{N-1} \ \leq \ {\alpha} \ \leq \  2^{N-1}  - 1
\end{equation}
Therefore, using two's complement, every such coefficient ${\alpha}$
can be encoded with $N$ bits. 
In addition, the integer $N$ is chosen such that $N$ writes
\begin{equation}
\label{eq:minimalConditionsForN}
N = K M  \ \ {\rm with} \ \ K \neq N \ \ {\rm and} \ \  M \neq N,
\end{equation}
for  $K, M \in {\N}$.

\ifnewversion
It is helpful to think of $M$ as a {\em small number}
in comparison to $K$ and $d$, say that $M$ is in the order of 
the bit-size, called $w$, of a machine word.
For the theoretical analysis 
of our algorithm, we shall assume, as in  Theorem~\ref{thrm:complexity},
that $K$ and $M$ are  
functions of $d$ satisfying $K \in {\Theta}(d)$ and
$M \in \Theta (\log{d})$.
We denote by $\mbox{{\sf DetermineBase}}(a,b,w)$
a function call returning $(N, K, M)$ satisfying those 
constraints as well as Relation~(\ref{eq:minimalConditionsForN}).
\fi

\iflongversion
\paragraph{From ${\Z}[y]$ to ${\Z}[x,y]$.}
\fi
Let $(N, K, M) := \mbox{{\sf DetermineBase}}(a,b,w)$
and define ${\beta} =  2^M$.
We write
\begin{equation}
a_i = \sum_{j=0}^{K-1} \, a_{i,j} {\beta}^j, \ \  {\rm and} \ \ 
b_i = \sum_{j=0}^{K-1} \, b_{i,j} {\beta}^j,
\end{equation}
where each $a_{i,j}$ and $b_{i,j}$ are signed integers
in the closed range $[-2^{M - 1}, 2^{M - 1}  - 1]$.
Then, we define 
\begin{equation}
A(x,y) = \sum_{i=0}^{d-1} \, (\sum_{j=0}^{K-1} \, a_{i,j} x^j) y^i, 
B(x,y) = \sum_{i=0}^{d-1} \, (\sum_{j=0}^{K-1} \, b_{i,j} x^j) y^i,
\end{equation}
and
\begin{equation}
\begin{array}{l}
C(x,y) := A(x,y) B(x,y) \ \ {\rm with}  \\
C(x,y) = \sum_{i=0}^{2d-2} \, \left(\sum_{j=0}^{2K-2} \, c_{i,j} x^j\right) y^i, \\
\end{array}
\end{equation}
where $c_{i,j} \in {\Z}$.
By ${\sf BivariateRepresentation}(a,N,K,M)$, 
we denote a function call returning $A(x,y)$ as defined above.
Observe that the polynomial $c(y)$ is clearly
recoverable from $C(x,y)$ 
\iflongversion
as
\begin{equation}
\begin{array}{rcll}
C({\beta},y) & =  & A({\beta},y) B({\beta},y) \\
             & =  & a(y) b(y) \\
             & = &  c(y).
\end{array}
\end{equation}
\else
by evaluating this latter polynomial at 
$x = {\beta}$.
\fi

The following sequence of equalities
will be useful:
\iflongversion
\begin{equation}
\label{eq:sequenceOfEqualities}
\begin{array}{rcl}
C(x,y) & = & A(x,y) \, B(x,y) \\
       & = & \left( \sum_{i=0}^{d-1} \, (\sum_{j=0}^{K-1} \, a_{i,j} x^j) y^i  \right)
              \left( \sum_{i=0}^{d-1} \, (\sum_{j=0}^{K-1} \, b_{i,j} x^j) y^i  \right) \\
      & = &  \sum_{i=0}^{2d-2} 
              \left(  \sum_{{\ell} + m = i} \left( \sum_{k=0}^{K-1} a_{{\ell},k} x^k \right)
                                         \left( \sum_{h=0}^{K-1} b_{m,h} x^h \right)
              \right) y^i \\ 
      & = &  \sum_{i=0}^{2d-2} 
              \left(  \sum_{{\ell} + m = i} 
               \left( \sum_{j=0}^{2K-2} 
                \left(       \sum_{k+h=j}  a_{{\ell},k}  b_{m,h} \right) x^j
                \right)  \right)   y^i \\ 
\iflongversion
      & = &  \sum_{i=0}^{2d-2} 
               \left(  \sum_{j=0}^{2K-2} 
              c_{i,j} x^j \right)   y^i \\ 
\fi
      & = &   \sum_{j=0}^{2K-2}  
               \left(  \sum_{i=0}^{2d-2} 
              c_{i,j}  y^i \right) x^j  \\ 

      & = &  \sum_{j=0}^{K-1}  \left( \sum_{i=0}^{2d-2}  c_{i,j} y^i \right) x^j
             \ + \ 
          x^K  \sum_{j=0}^{K-2} \left( \sum_{i=0}^{2d-2}  c_{i,j+K} \; y^i \right) x^j,
\end{array}
\end{equation}
\else
\begin{equation}
\label{eq:sequenceOfEqualities}
\small
\begin{array}{rl}
C  = & A\, B \\
        = & \left( \sum_{i=0}^{d-1} \, (\sum_{j=0}^{K-1} \, a_{i,j} x^j) y^i  \right)
              \left( \sum_{i=0}^{d-1} \, (\sum_{j=0}^{K-1} \, b_{i,j} x^j) y^i  \right) \\
       = &  \sum_{i=0}^{2d-2} 
              \left(  \sum_{{\ell} + m = i} \left( \sum_{k=0}^{K-1} a_{{\ell},k} x^k \right)
                                         \left( \sum_{h=0}^{K-1} b_{m,h} x^h \right)
              \right) y^i \\ 
       = &  \sum_{i=0}^{2d-2} 
              \left(  \sum_{{\ell} + m = i} 
               \left( \sum_{j=0}^{2K-2} 
                \left(       \sum_{k+h=j}  a_{{\ell},k}  b_{m,h} \right) x^j
                \right)  \right)   y^i \\ 
       = &  \sum_{i=0}^{2d-2} 
               \left(  \sum_{j=0}^{2K-2} 
              c_{i,j} x^j \right)   y^i \\ 
       = &   \sum_{j=0}^{2K-2}  
               \left(  \sum_{i=0}^{2d-2} 
              c_{i,j}  y^i \right) x^j  \\ 
       = &  \sum_{j=0}^{K-1}  \left( \sum_{i=0}^{2d-2}  c_{i,j} y^i \right) x^j \\
         & 
           \  + \
          x^K  \sum_{j=0}^{K-2} \left( \sum_{i=0}^{2d-2}  c_{i,j+K} \; y^i \right) x^j, \\
\end{array}
\end{equation}
\fi
\ifnewversion
where we have
\begin{equation}
c_{i,j}  =  \sum_{{\ell} + m = i}  \sum_{k+h=j}  a_{{\ell},k}  b_{m,h},
0 \leq i \leq 2d-2, 0 \leq j \leq 2K-2,
\end{equation}
\correction{1}{with the convention
\begin{equation}
c_{i,2K-1} \ := \ 0 \ \ {\rm for} \ \ 0 \leq i \leq 2d-2.
\end{equation}
}
\fi
Since the modular products
$A(x,y) B(x,y) \mod{\langle x^K + 1 \rangle}$
and
$A(x,y) B(x,y) \mod{\langle x^K - 1 \rangle}$
are of interest, we define the bivariate polynomial over {\Z}
\iflongversion
\begin{equation}
C^{+}(x,y) \ := \ \sum_{i=0}^{2d-2} \, c^{+}_i(x) \, y^i 
 \ \ {\rm where} \ \ 
c^{+}_i(x) := \sum_{j=0}^{K-1} \, c^{+}_{i,j} \,  x^j
 \ \ {\rm and} \ \ 
c^{+}_{i,j} := c_{i,j} - c_{i, j+ K}
\end{equation}
\else
\begin{equation}
C^{+}(x,y) \ := \ \sum_{i=0}^{2d-2} \, c^{+}_i(x) \, y^i 
 \ \ {\rm where} \ \ 
c^{+}_i(x) := \sum_{j=0}^{K-1} \, c^{+}_{i,j} \,  x^j
\end{equation}
with $c^{+}_{i,j} := c_{i,j} - c_{i, j+ K}$,
\fi
and the bivariate polynomial over {\Z}
\iflongversion
\begin{equation}
\label{eq:Cminus}
C^{-}(x,y) \ := \ \sum_{i=0}^{2d-2} \, c^{-}_i(x) \, y^i 
 \ \ {\rm where} \ \ 
c^{-}_i(x) := \sum_{j=0}^{K-1} \, c^{-}_{i,j} \,  x^j
 \ \ {\rm and} \ \ 
c^{-}_{i,j} := c_{i,j} + c_{i, j+ K}.
\end{equation}
\else
\begin{equation}
\label{eq:Cplus}
C^{-}(x,y) \ := \ \sum_{i=0}^{2d-2} \, c^{-}_i(x) \, y^i 
 \ \ {\rm where} \ \ 
c^{-}_i(x) := \sum_{j=0}^{K-1} \, c^{-}_{i,j} \,  x^j
\end{equation}
with $c^{-}_{i,j} := c_{i,j} + c_{i, j+ K}.$
\fi
Thanks to Equation~(\ref{eq:sequenceOfEqualities}),
we observe that we have
\begin{equation}
\label{eq:CplusandCminus}
\begin{array} {rcl}
C^{+}(x,y) & \equiv & A(x,y) B(x,y) \mod{\langle x^K + 1\rangle}, \\
C^{-}(x,y) & \equiv & A(x,y) B(x,y) \mod{\langle x^K - 1\rangle}. \\
\end{array}
\end{equation}
Since the polynomials $x^K + 1$ and $x^K - 1$ are coprime 
for the integer $K \geq 1$,
we deduce Equation (\ref{eq:recoveringC}).

Since ${\beta}$ is a power of $2$, evaluating the polynomials
$C^{+}(x,y)$, $C^{-}(x,y)$ and thus $C(x,y)$ (whose coefficients are
assumed to be in binary representation) at $x = {\beta}$ amounts only
to {\em addition} and {\em shift} operations.  A precise algorithm is
described in Section~\ref{subsec:recoveringcy}.
Before that, we turn our attention to computing
$C^{+}(x,y)$ and $C^{-}(x,y)$ via FFT techniques.

\iflongversion
\paragraph{From ${\Z}[x,y]$ to ${\Z}/m[x,y]$.}
\else
\subsection{Computing $C^{+}(x,y)$ and $C^{-}(x,y)$ via FFT}
\label{subsec:recoveringcy}
\fi

From Equation~(\ref{eq:CplusandCminus}), 
it is natural to consider using FFT techniques
for computing both $C^{+}(x,y)$ and $C^{-}(x,y)$.
Thus, in order to compute over a finite ring supporting FFT,
we estimate the size of the coefficients of $C^{+}(x,y)$ and $C^{-}(x,y)$.
Recall that for $0 \leq i \leq 2d-2$, we have
\iflongversion
\begin{equation}
\begin{array}{rcll}
c^{+}_{i,j} & = & c_{i,j} - c_{i,j+K} & \\
          & = & \sum_{{\ell} + m = i}  \sum_{k+h=j}  a_{{\ell},k}  b_{m,h} -
                \sum_{{\ell} + m = i}  \sum_{k+h=j+K}  a_{{\ell},k}  b_{m,h}. \\
\end{array}
\end{equation}
\else
\begin{equation}
\begin{array}{rcll}
c^{+}_{i,j} & = & c_{i,j} - c_{i,j+K} & \\
          & = & \sum_{{\ell} + m = i}  \sum_{k+h=j}  a_{{\ell},k}  b_{m,h} \\
          &   &
                -  \sum_{{\ell} + m = i}  \sum_{k+h=j+K}  a_{{\ell},k}  b_{m,h}. \\
\end{array}
\end{equation}
\fi
Since each $a_{{\ell},k}$ and each $ b_{m,h}$ has bit-size at most $M$,
the absolute value of 
each coefficient $c^{+}_{i,j}$ is bounded over by $2 \, d \, K \, 2^{2 M}$.
The same holds for the coefficients $c^{-}_{i,j}$.

\ifnewversion
Since the coefficients $c^{+}_{i,j}$ and $c^{-}_{i,j}$ may be negative,
we consider a prime number $p$ such that we have
\begin{equation}
\label{eq:recoveryModulo}
p > 4 \,d \, K \, 2^{2 M}.
\end{equation}
From now on, depending on the context,
we freely view the coefficients $c^{+}_{i,j}$ and $c^{-}_{i,j}$ 
either as elements of {\Z} or as elements of ${\Z}/p$.
Indeed, the integer $p$ is large enough for this identification
and we use the integer interval $[- \frac{p-1}{2}, \frac{p-1}{2}]$
to represent the elements of ${\Z}/p$.

The fact that we follow a {\em big prime approach},
instead of an approach using machine word size
primes and the Chinese Remaindering Algorithm,
will be justified in Section~\ref{sec:cmmxxmultiplication-complexity}.

A second requirement for the prime number $p$ 
is that the field ${\Z}/p$ should admit appropriate
primitive roots of unity for computing the polynomials 
$C^{+}(x,y)$ and $C^{-}(x,y)$ via cyclic convolution 
and negacylic convolution as in Relation (\ref{eq:CplusandCminus}),
that is, both $2 d -1$ and $K$ must divide $p - 1$.
In view of utilizing 2-way FFTs, if $p$ writes 
$2^k q + 1$ for an integer $q$, we must have:
\begin{equation}
\label{eq:HavingAppropriatePrimitiveRootsOfUnity}
(2 d -1) \leq 2^k \ \ {\rm and} \ \
K \leq 2^k.
\end{equation}
It is well-known that there are approximately
$\frac{h}{2^{k-1} \, \log{(h)}}$ prime numbers
$p$ of the form $2^k q + 1$\correction{1}{for a positive integer $q$}
and such that $p < h$ holds, see~\cite{Gathen:2003:MCA:945759}.
Hence, the number of prime numbers 
of the form $2^k q + 1$ less than $2^{\ell}$
and greater than or equal to $2^{\ell -1}$ is approximately
$\frac{2^{\ell -1}}{2^{k-1} \, {\ell}}$.
For this latter fraction to be at least one, we must have
\begin{equation*}
2^{{\ell} - \log_2({\ell}) + 1}  \geq 2^{k}.
\end{equation*}
With (\ref{eq:HavingAppropriatePrimitiveRootsOfUnity})
this yields:
\begin{equation*}
2^{{\ell} - \log_2({\ell}) + 1}  \geq (2 d -1) \ \ {\rm and} \ \
2^{{\ell} - \log_2({\ell}) + 1}  \geq K,
\end{equation*}
\correction{1}{from where we derive the following relation for 
choosing ${\ell}$:}
\begin{equation}
\label{eq:HavingAppropriatePrimitiveRootsOfUnity2}
{\ell} - \log_2({\ell}) \ \geq \ {\max}(\log_2(d), \log_2(K)).
\end{equation}

We denote by ${\sf RecoveryPrime}(d, K, M)$
a function call returning 
a prime number $p$ satisfying Relation (\ref{eq:recoveryModulo})
and (\ref{eq:HavingAppropriatePrimitiveRootsOfUnity2}).
\ifcomment
We shall see in Section~\ref{sec:cmmxxmultiplication-complexity}
that under two realistic assumptions, namely
$M \in \Theta(d)$ and $K \in {\Theta}(d)$,
Relation (\ref{eq:recoveryModulo})
implies Relation (\ref{eq:HavingAppropriatePrimitiveRootsOfUnity2}).
\fi
Finally, we denote by $e$ the smallest number of $w$-bit
words necessary to write $p$. Hence, we have
\begin{equation}
\label{eq:MinimumNumberOfMachineWords}
e \ \geq \ \left\lceil 
           \frac{2 + \lceil {\log}_2(d \, K) \rceil + 2M}{w} 
             \right\rceil.
\end{equation}

\iflongversion
\paragraph{Cyclic and negacylic convolutions in ${\Z}/p[x,y]$.}
\fi
Let ${\theta}$ be a $2 K$-th primitive root
of unity in ${\Z}/p$.
We define $\omega = {\theta}^2$; thus, $\omega$
is a $K$-th primitive root in ${\Z}/p$.
For univariate polynomials $u(x), v(x) \in {\Z}[x]$ of degree
at most $K-1$,
computing $u(x) \, v(x) \mod{\langle  x^K - 1, p\rangle}$
via FFT is a well-known operation, see Algorithm 8.16
in~\cite{Gathen:2003:MCA:945759}.
Using the {\em row-column} algorithm for two-dimensional FFT, one can compute
$C^{-}(x,y) \, \equiv \, A(x,y) B(x,y) \mod{\langle  x^K - 1, p\rangle}$,
see~\cite{DBLP:journals/ijfcs/MazaX11,DBLP:conf/hpcs/MazaX09} for details.
We denote by ${\sf CyclicConvolution}(A, B, K, p)$ the result of this
calculation.

We turn our attention to the negacylic convolution, 
namely $A(x,y) B(x,y) \mod{\langle  x^K + 1, p\rangle}$.
\iflongversion
We observe that the following holds:
\begin{equation}
C^{+}(x,y) \ \equiv \ A(x,y) B(x,y) \mod{\langle x^K + 1, p \rangle}
\ \ \iff \ \ 
C^{+}({\theta} x,y) \ \equiv \ A({\theta} x,y) B({\theta} x,y) \mod{\langle x^K - 1, p \rangle}
\end{equation}
\else
We observe that
$C^{+}(x,y) \ \equiv \ A(x,y) B(x,y) \mod{\langle x^K + 1, p \rangle}$
holds if only if
$C^{+}({\theta} x,y) \ \equiv \ A({\theta} x,y) B({\theta} x,y) \mod{\langle x^K - 1, p \rangle}$
does.
\fi
\iflongversion
Thus, defining 
\begin{equation}
C'(x,y) := C^{+}({\theta}x,y), \ \ 
A'(x,y) := A({\theta} x,y) \ \ {\rm and} \ \ 
B'(x,y) := B({\theta} x,y), 
\end{equation}
\else
Thus, defining 
$C'(x,$ $y) := C^{+}({\theta}x,y)$,
$A'(x,y) := A({\theta} x,y)$ and
$B'(x,y) := B({\theta} x,y)$,
\fi
we are led to compute 
\iflongversion
\begin{equation}
A'(x,y) B'(x,y) \mod{\langle x^K - 1, p \rangle},
\end{equation}
\else
$A'(x,y) B'(x,y) \mod{\langle x^K - 1, p \rangle},$
\fi
which can be done as ${\sf CyclicConvolution}(A', B', K, p)$.
Then, the polynomial $C^{+}(x,y) \mod{\langle x^K - 1, p \rangle}$
is recovered from $C'(x,y) \mod{\langle x^K - 1, p \rangle }$ as
\begin{equation}
C^{+}(x,y)   \ \equiv \  C'({\theta}^{-1}x,y) \mod{\langle x^K - 1, p \rangle},
\end{equation}
and we denote by  ${\sf NegacyclicConvolution}(A, B, K, p)$ 
the result of this process.
We dedicate the following section to the final step of our algorithm,
that is, the recovery of the product $c(y)$ from the polynomials 
$C^{+}(x,y)$ and $C^{-}(x,y)$.

\fi

\ifnewversion
\subsection{Recovering $c(y)$ from $C^{+}(x,y)$ and $C^{-}(x,y)$}
\label{subsec:recoveringcy}

We naturally assume that all numerical coefficients are stored in
binary representation.  Thus, recovering $c(y)$ as $C({\beta},y)$ from
Equation (\ref{eq:recoveringC}) involves only addition and shift
operations.  Indeed, ${\beta}$ is a power of $2$.  Hence, the
algebraic complexity of this recovery is essentially proportional to
the sum of the bit sizes of $C^{+}(x,y)$ and $C^{-}(x,y)$.  Therefore,
the arithmetic count for computing these latter polynomials (by means
of cyclic and negacyclic convolutions) dominates that of recovering
$c(y)$.
Nevertheless, when implemented on a modern computer hardware, this
recovery step may contribute in a significant way to the total running
time.  The reasons are that: (1) both the convolution computation and
recovery steps incur similar amounts of cache misses, and (2) the
memory traffic implied by those cache misses is a significant portion
of the total running time.

\ifcomment
Indeed, the ratio of its arithmetic count to its cache misses
is in the same order of magnitude as the cache line size, that we denote by $L$.
Similarly, for the cyclic and negacyclic convolution computations,
which are based on FFTs, the same ratio is in the order of $L$,
see the paper~\cite{DBLP:journals/talg/FrigoLPR12}.
Since $L$ is generally much less than the ratio between
the number of clock cycles for a machine word operation 
and the number of clock cycles for bringing a cache line
from the main memory to the cache, it follows that 
cache misses contribute in a significant manner
to the timings of both the convolution computations and recovery steps.
Consequently, the fact that operations incur the same amount of cache misses
explains why the recovery step may contribute in a significant way
to the total running time.
\fi

We denote by ${\sf RecoveringProduct}(C^{+}(x,y), C^{-}(x,y), {\beta})$
a function call recovering $c(y)$
from $C^{+}(x,y)$, $C^{-}(x,y)$ and ${\beta} = 2^M$.
We start by stating below a simple procedure for this operation:
\medskip
\begin{enumerateshort}
\item $u(y)$ := $C^{+}({\beta},y)$,
\item $v(y)$ := $C^{-}({\beta},y)$,
\item $c(y)$ := $\frac{u(y) + v(y)}{2} + \frac{-u(y) + v(y)}{2} \, 2^N$.
\end{enumerateshort}
\medskip
\medskip
To further describe this operation and, later on, in order to discuss
its cache complexity and parallelization, we specify the data layout.
From Relation (\ref{eq:MinimumNumberOfMachineWords}), we can assume 
 that each coefficient of the bivariate polynomials $C^{+}(x,y)$,  $C^{-}(x,y)$
can be encoded within $e$ machine words.
Thus, we assume that $C^{+}(x,y)$ (resp. $C^{-}(x,y)$)
is represented by an array of $(2d-1) \, K \, e$ machine words
such that, for $0 \leq i \leq 2d-2$ and $0 \leq j \leq K-1$, 
the coefficient  $c^{+}_{i,j}$ (resp. $c^{-}_{i,j}$)
is written between the positions $(K \, i + j) e$
and $(K \, i + j) e + e-1$, inclusively.
Thus, this array can be regarded as the encoding of a
2-D matrix whose $i$-th row is $c^{+}_i(x)$ (resp.  $c^{-}_i(x)$).
Now, we write 
\begin{equation}
u(y) := \sum_{i=0}^{2d-2} \, u_i y^i \ \ {\rm and} \ \ 
v(y) := \sum_{i=0}^{2d-2} \, v_i y^i;
\end{equation}
thus, from the definition of $u(y)$, $v(y)$, 
for $0 \leq i \leq 2d-2$, we have
\begin{equation}
\label{eq:uividefinition}
u_i =  \sum_{j=0}^{K-1} \, c^{+}_{i,j} \, 2^{M \, j} \ \ {\rm and} \ \ 
v_i =  \sum_{j=0}^{K-1} \, c^{-}_{i,j} \, 2^{M \, j}.
\end{equation}
Denoting by $H^{+}$, $H^{-}$ the largest absolute value of
a coefficient in $C^{+}(x,y)$, $C^{-}(x,y)$, we deduce 
\begin{equation}
\label{eq:uivi}
| u_i | \ \leq \ H^{+} {\frac { \left(  \left( {2}^{M} \right) ^{K}-1 \right) }{{2}^{M}-1}}
\ \ {\rm and} \ \ 
| v_i | \ \leq \ H^{-} {\frac { \left(  \left( {2}^{M} \right) ^{K}-1 \right) }{{2}^{M}-1}}.
\end{equation}
From the discussion justifying Relation~(\ref{eq:recoveryModulo}),
we have
\begin{equation}
H^{+}, \, H^{-} \ \leq \  2 \, d \, K \, 2^{2 M},
\end{equation}
and with~(\ref{eq:uivi}) we derive
\begin{equation}
\begin{array}{rcl}
| u_i |, \, | v_i | 
                   & \leq & 2 \, d \, K \, 2^{M + N}. \\
\end{array}
\end{equation}
Indeed, recall that $N = K \, M$ holds.
We return to the question of data layout.
Since each of $c^{+}_{i,j}$ or $ c^{-}_{i,j}$ is a signed integer
fitting within $e$ machine words,  it follows from~(\ref{eq:uivi}) that 
each of the coefficients $u_i, v_i$ can be encoded within 
\begin{equation}
f \ := \ \lceil N/w \rceil + e
\end{equation}
machine words. Hence, we store each of the polynomials $u(y), v(y)$ in an array
of $(2d-1) \times f$ machine words such that
the coefficient in degree $i$ is located between position 
$f \, i$ and position $f \, (i + 1) -1$.
Finally, we come to the computation of $c(y)$.
We have
\begin{equation}
c_i \ = \ \frac{u_i + v_i}{2} \ +  \  2^N \ \frac{v_i - u_i }{2}, 
\end{equation}
which implies 
\begin{equation}
\label{eq:cibigbound}
| c_i | \ \leq \ 2 \, d \, K \, 2^{M + N} (1 + 2^N).
\end{equation}
Relation~(\ref{eq:cibigbound}) implies  that
the polynomial $c(y)$ can be stored within an array
of $(2d-1) \times 2f$ machine words.
Of course, a better bound than (\ref{eq:cibigbound})
can be derived by simply expanding
the product $a(y) \, b(y)$, leading to
\begin{equation}
\label{eq:citighterbound}
| c_i | \ \leq \ d \, 2^{2 N - 2}.
\end{equation}
The ratio between the two bounds given by Relations~(\ref{eq:cibigbound})
and (\ref{eq:citighterbound}) tells us that the extra amount
of space required by our algorithm is $O({\log}_2(K) + M)$ bits
per coefficient of $c(y)$.
In practice, we aim at choosing $K, M$ such that
$M \in {\Theta}({\log}_2(K))$, and if possible $M \leq w$.
Hence, this extra space amount can be regarded as small
and thus satisfactory.

\fi

\ifnewversion
\subsection{The algorithm in pseudo-code}
\label{subsec:pseudocode}

With the procedures that were defined in this section,
we are ready to state our algorithm for integer
polynomial multiplication.

\medskip
\begin{enumerateshort}
\item[] {\bf Input:} $a(y),b(y) \in {\Z}[y]$.
\item[] {\bf Output:} the product $a(y) \, b(y)$
\item[ {\bf 1:} ] $(N, K, M)$ := \mbox{${\sf DetermineBase}(a(y),b(y),w)$}
\item[ {\bf 2:} ]   $A(x,y)$ := \mbox{${\sf BivariateRepresentation}(a(y),N,K,M)$}
\item[ {\bf 3:} ]   $B(x,y)$ := \mbox{${\sf BivariateRepresentation}(b(y),N,K,M)$}
\item[ {\bf 4:} ]   $p$ := \mbox{${\sf RecoveryPrime}(d, K, M)$}

\item[ {\bf 5:} ]   $C^{-}(x,y)$ := \mbox{${\sf CyclicConvolution}(A, B, K, p)$}
\item[ {\bf 6:} ]   $C^{+}(x,y)$ := \mbox{${\sf NegacyclicConvolution}(A, B, K, p)$}
\item[ {\bf 7:} ]   $c(y)$ := \mbox{${\sf RecoveringProduct}(C^{+}(x,y), C^{-}(x,y), 2^M)$}
\item[ {\bf 8:} ]    {\bf return} $c(y)$
\end{enumerateshort}
\medskip

In order to analyze the complexity of our algorithm, it remains to
specify the data layout for $a(y)$, $b(y)$, $A(x,y)$, $B(x,y)$.  Note
that this data layout question was handled for $C^{-}(x,y)$,
$C^{+}(x,y)$ and $c(y)$ in Section~\ref{subsec:recoveringcy}.

In the sequel, we view $a(y)$, $b(y)$ as {\em dense} in the sense
that each of their coefficients is assumed to be essentially of 
the same size. Hence, from the definition of $N$, see Relation (\ref{eq:definingN}),
we assume that each of $a(y)$, $b(y)$ is stored within an array
of $d \times \lceil N/w \rceil$ machine words such that
the coefficient in degree $i$ is located between positions 
$\lceil N/w \rceil i$ and $\lceil N/w \rceil (i + 1) -1$.

Finally, we assume that each of the bivariate integer polynomials $A(x,y)$,  $B(x,y)$
is represented by an array of $d \times K$ machine words
whose $(K \times i + j)$-th coefficient is $a_{i,j}$, $b_{i,j}$
respectively, for $0 \leq i \leq d-1$ and $0 \leq j \leq K-1$.

\fi

\ifnewversion
\subsection{Parallelization}
\label{subsec:parallelization}

One of the initial motivations of our algorithm design is to take
advantage of the parallel FFT-based routines for multiplying dense
multivariate polynomials over finite fields that have been proposed
in~\cite{DBLP:conf/hpcs/MazaX09,DBLP:journals/ijfcs/MazaX11}.
To be precise, these routines provide us
with a parallel implementation of the procedure
{\sf CyclicConvolution}, from which we easily
derive a parallel implementation of {\sf NegacyclicConvolution}.

Lines {\bf 1} and {\bf 4} can be ignored in the analysis
of the algorithm. Indeed, 
\iflongversion
as we shall explain in 
Section~\ref{sec:cmmxxmultiplication-implementation},
\fi
one can simply implement {\sf DetermineBase} and {\sf RecoveryPrimes}
by looking up in precomputed tables.
\ifthesis 
For instance, Table~\ref{table:goodN-2primes} in Appendix~\ref{appendix:table}
is being used for determining the base using 2 primes.
\fi

For parallelizing Lines {\bf 2} and {\bf 3}, it is sufficient in practice
to convert concurrently all the coefficients of $a(y)$ and $b(y)$
to univariate polynomials of ${\Z}[y]$.
Similarly, for parallelizing Line {\bf 7}, 
it is  sufficient again to compute concurrently
the coefficients of $u(y)$, $v(y)$ and then those of $c(y)$.

\fi

\section{Complexity analysis}
\label{sec:cmmxxmultiplication-complexity}

In this section, we analyze the algorithm stated in
Section~\ref{subsec:pseudocode}.  We estimate its {\em work} and {\em
  span} as defined in the {\em fork-join concurrency
  model}
  \ifthesis 
  introduced in Section~\ref{sec::fork-join}.
  \else
  ~\cite{DBLP:journals/siamcomp/BlumofeL98}.  The reader
unfamiliar with this model can regard the work as the time complexity
on a multitape Turing machine~\cite{DBLP:books/daglib/0080802} and the
span as the minimum parallel running time of a ``fork-join program''.
Such programs use only two primitive language constructs, namely {\tt
  fork} and {\tt join}, in order to express concurrency.
  \fi
Since the fork-join model has no primitive constructs for defining
parallel for-loops, each of those loops is simulated by a
divide-and-conquer procedure for which non-terminal recursive calls
are forked, see ~\cite{DBLP:journals/tjs/Leiserson10} for details.
Hence, in the fork-join model, the bit-wise comparison of two vectors
of size $n$ has a span of $O({\log}(n))$ bit operations.  This is
actually the same time estimate as in the
Exclusive-Read-Exclusive-Write
PRAM~\cite{DBLP:journals/siamcomp/StockmeyerV84,Gibbons:1989:MPP:72935.72953}
model, but for a different reason.

We shall also estimate the {\em cache
  complexity}~\cite{DBLP:journals/talg/FrigoLPR12} of the serial
counterpart of our algorithm for an ideal cache of $Z$ words and with
$L$ words per cache line.
\ifthesis
Note that the ratio work to cache complexity indicates how an
algorithm is capable of re-using cached data. Hence the larger is the
ratio, the better the algorithm is.
\else
The reader unfamiliar with this notion may understand it as a measure
of memory traffic or, equivalently on multicore processors, as a measure
of data communication.  Moreover, the reader should note that the ratio work
to cache complexity indicates how an algorithm is capable of re-using
cached data. Hence, the larger is the ratio, the better the algorithm is.  \fi

\ifnewversion
We denote by $W_i$, $S_i$, $Q_i$ the work, span and cache complexity
of Line ${\bf i}$ in the algorithm stated in
Section~\ref{subsec:pseudocode}.  As mentioned before, we can ignore
the costs of Lines {\bf 1} and {\bf 4}.  Moreover, we can use $ W_2$,
$S_2$, $Q_2$ as estimates for $ W_3$, $S_3$, $Q_3$, respectively.
Similarly, we can use the estimates of Line {\bf 5} for the costs of
Line {\bf 6}.  Thus, we only analyze the costs of Lines {\bf 2}, {\bf
5} and {\bf 7}.

\ifnewversion
\subsection{Choosing $K$, $M$ and $p$}
\label{sec:choosingKMp}

In order to analyze the costs associated with the cyclic and
negcyclic convolutions, we shall specify how $K$, $M$, $p$ are
chosen. We shall assume thereafter
that $K$ and $M$ are  
functions of $d$ satisfying the following asymptotic
relations:
\begin{equation}
\label{eq:KMd}
K \in {\Theta}(d) \ \  {\rm and} \ \ 
M \in {\Theta} (\log{d}).
\end{equation}
It is a routine exercise to check that 
these assumptions together with 
Relation (\ref{eq:recoveryModulo})
imply Relation (\ref{eq:HavingAppropriatePrimitiveRootsOfUnity2}).

Relations (\ref{eq:KMd}) and (\ref{eq:MinimumNumberOfMachineWords})
imply that we can choose $p$ and thus $e$ such that we have
\begin{equation}
\label{eq:MinimumNumberOfMachineWords2}
e \in {\Theta} (\log{d}).
\end{equation}

Here comes our most important assumption:
one can choose $p$ and thus $e$ such that  computing 
an FFT of  a vector of size $s$ over ${\Z}/p[x]$,
amounts to
\begin{equation}
\label{eq:FurerHypothesis}
{\sf F}_{\rm arith}(e,s) \ \in \ O \left(s \frac{\log(s)}{\log(e)}\right)
\end{equation}
arithmetic operations in ${\Z}/p$, whenever $e \in {\Theta}
(\log{s})$ holds.
Since each arithmetic operation in ${\Z}/p$
can be done within $O(e \log(e) \log\log(e))$ machine-word operations
(using the multiplication  algorithm of Sch{\"o}nhage and Strassen).
Hence:
\begin{equation}
\label{eq:Fes}
{\sf F}_{\rm word}(e,s) \ \in \ O(s \, e \, \log(s) \, \log\log(e))
\end{equation}
machine-word operations, whenever $e \in {\Theta} (\log{s})$ holds.
Using the fork-join model, the corresponding span
is $O(\log^2(s) \log^2(e) \log\log(e))$ machine-word operations.

\ifrevision

\correction{1}{
Relation~(\ref{eq:FurerHypothesis}) can be derived 
from~\cite{DBLP:journals/corr/CovanovT15}
assuming that $p$ is a {\em generalized Fermat prime}.
Table~\ref{tab:SparseRadixGeneralizedFermatNumbers}
lists generalized Fermat primes of practical interest.
Moreover, by adapting the results and proof strategy of~\cite{Harvey2016}
to the analysis in Section 2.6 of~\cite{Covanov2014}, we can obtain a similar
binary complexity as that of 
Relation~(\ref{eq:FurerHypothesis})
for a prime $p$ which is not necessarily a generalized Fermat prime.
}
\else
Relation (\ref{eq:FurerHypothesis}) can be derived 
with other technical assumptions about ${\Z}/p$ from
the use of the algorithm proposed in~\cite{Harvey2016}.
The analysis proposed in Section 2.6 of~\cite{Covanov2014}
proposes an analysis specific to this case.
Later, Relation (\ref{eq:FurerHypothesis}) was derived
in~\cite{DBLP:journals/corr/CovanovT15}
assuming that $p$ is a {\em generalized Fermat prime}.
Table~\ref{tab:SparseRadixGeneralizedFermatNumbers}
lists generalized Fermat primes of practical interest.
See~\cite{Covanov2014} for more details.
\fi

\begin{table}[htbp]
\centering
\begin{tabular}{|c|c|} \hline 
$p$ & ${\max}\{ 2^{k} \ {\rm s.t.} \ 2^k \, \mid \,  p-1 \}$ \\ \hline   
$(2^{63} + 2^{53})^2 + 1$ & $2^{106}$ \\ \hline 
$(2^{64} - 2^{50})^4 + 1$ & $2^{200}$ \\ \hline 
$(2^{63} + 2^{34})^8 + 1$ & $2^{272}$ \\ \hline 
$(2^{62} + 2^{36})^{16} + 1$ & $2^{576}$ \\ \hline  
$(2^{62} + 2^{56})^{32} + 1$ & $2^{1792}$ \\ \hline 
$(2^{63} - 2^{40})^{64} + 1$ & $2^{2500}$ \\ \hline 
$(2^{64} - 2^{28})^{128} + 1$ & $2^{3584}$ \\ \hline 
\end{tabular}
\caption{\small Generalized Fermat primes of practical interest}
\label{tab:SparseRadixGeneralizedFermatNumbers}
\end{table}
\vspace{-2em}

\fi

\iflongversion
\paragraph{Analysis of \mbox{${\sf BivariateRepresentation}(a(y),N,K,M)$}.}
\else
\ifcorr
\subsection{Analysis of \\ \mbox{${\sf BivariateRepresentation}(a(y),N,K,M)$}}
\else
\subsection{Analysis of \mbox{${\sf BivariateRepresentation}(a(y),N,K,M)$}}
\fi
\fi
Converting each coefficient of $a(y)$ to a univariate polynomial of ${\Z}[x]$
requires $O(N)$ bit operations;
thus,
\begin{equation}
\label{eq:WTwo}
W_2 \in O(d \, N) \ \  {\rm and}  \ \ S_2 \in O({\log}(d) \, N).
\end{equation}
The latter ${\log}(d)$ factor comes from the fact that the
parallel for-loop corresponding to ``for each coefficient of $a(y)$''
is executed as a recursive function with $O({\log}(d))$ nested
recursive calls.
Considering the cache complexity and taking into account the data
layout specified in Section~\ref{subsec:pseudocode}, one can observe that
converting $a(y)$ to $A(x,y)$ leads to $O(\lceil \frac{d N}{w L}\rceil
+ 1)$ cache misses for reading $a(y)$ and $O(\lceil \frac{d K
e}{L}\rceil + 1)$ cache misses for writing $A(x,y)$.  
Therefore, we have:
\begin{equation}
Q_2  \in O \left(\left\lceil \frac{d N}{w L} \right\rceil + 
            \left\lceil \frac{d K \log_2(d K)}{w L} \right\rceil + 1 \right).
\end{equation}

\ifnewversion
\subsection{Analysis of \mbox{${\sf CyclicConvolution}(A, B, K, p)$}}

Following the techniques
developed in~\cite{DBLP:conf/hpcs/MazaX09,DBLP:journals/ijfcs/MazaX11},
we compute $A(x,y) B(x,y) \mod{\langle x^K - 1, p \rangle}$
by 2-D FFT of format $K \times 2 d$.
In the direction of $y$, we need to do $K$ FFTs of size $2 d$
leading to $O(K \, F_{\rm word}(e,2 d))$ machine word operations.
In the direction of $x$, we need to compute $2 d$
 convolutions 
(i.e. products in ${\Z}[x] / \langle x^K -  1, p \rangle$)
leading to $O(2 d \, F_{\rm word}(e,K))$ machine word operations.
Using Relations (\ref{eq:KMd}), 
(\ref{eq:MinimumNumberOfMachineWords2}) and  (\ref{eq:Fes}), the work 
of one  2-D FFT of format $K \times 2 d$ amounts to:
\begin{equation}
\label{eq:WFive}
\begin{array}{rl}
O(K  F_{\rm word}(e,2 d)) + O(2 d  F_{\rm word}(e,K)) & = \\
O(K e 2 d (\log(2d) + \log(K)) \log\log(e)) & = \\
O(K e 2 d \log(2d K) \log\log(e))  & = \\
O(K M 2 d \log(2d K) \log\log(\log(d))) 
\end{array}
\end{equation}
machine word operations, while the span  amounts to:
\begin{equation}
\begin{array}{l}
O((\log(K) \log^2(d) + \log(d) \log^2(K)) \log^2(e) \log\log(e))   \\
= O(\log(K) \log(d) \log(d K) \log^2(\log(d)) \log\log(\log(d))).
\end{array}
\end{equation}
Hence:
\begin{equation}
\label{eq:WFive}
W_5 \in O(d K M {\log}(d K) \log\log(\log(d))) \ \ {\rm and} \ \
\end{equation}
 $S_5 \in O(\log(K) \log(d) \log(d K) \log^2(\log(d)) \log\log(\log(d)))$
machine word operations.
Finally, from the results of~\cite{DBLP:journals/talg/FrigoLPR12} 
and assuming that $Z$ is large enough to accommodate a few
elements of ${\Z}/p$, we have
\begin{equation}
Q_6 \in O(1 + (d e K/L) (1 + {\log}_{Z}(d e K))).
\end{equation}

\else
\input{cyclic-convolution-analysis}
\fi

\ifnewversion
\iflongversion
\paragraph{Analysis of \mbox{${\sf RecoveringProduct}(C^{+}(x,y), C^{-}(x,y), 2^M)$}.}
\else
\ifcorr 
\subsection{Analysis of \\ \mbox{${\sf RecoveringProduct}(C^{+}(x,y), C^{-}(x,y), 2^M)$}}
\else
\subsection{Analysis of \mbox{${\sf RecoveringProduct}(C^{+}(x,y), C^{-}(x,y), 2^M)$}}
\fi
\fi

Converting each coefficient of $u(y)$ and $v(y)$ 
from the corresponding coefficients of $C^{+}(x,y)$ and $C^{-}(x,y)$
requires $O(K \, e)$ bit operations.
Then, computing each coefficient of $c(y)$ 
requires $O(N + e \,  w)$ bit operations.
Thus, we have
\begin{equation}
\label{eq:WSeven}
W_{7} \in O(d \, (K \, e + N + e)) \ \ {\rm and} \ \ 
S_{7}  \in O( {\log}(d) \, (K \, e + N + e))
\end{equation}
word operations.
Converting $C^{+}(x,y)$, $C^{-}(x,y)$ to $u(y)$, $v(y)$ leads to 
$O(\lceil d \, K \, e / L \rceil + 1)$
cache misses for reading $C^{+}(x,y)$, $C^{-}(x,y)$ 
and
$O(\lceil d  (N/w + e) / L  \rceil + 1)$
cache misses for writing $u(y)$, $v(y)$.
This second estimate holds also for
the total number of cache misses for 
computing  $c(y)$ from $u(y)$, $v(y)$.
Thus, we have
\iflongversion
\begin{equation}
Q_{7} \in O(\lceil d \, (K \, e + N + e) / L\rceil + 1).
\end{equation}
\else
$Q_{7} \in O(\lceil d \, (K \, e + N + e) / L\rceil + 1).$
\fi
We note that the quantity $K \, e + N + e$ can be replaced
in above asymptotic upper bounds by $K ({\log}_2(d \, K) + 3 \, M)$.

\else
\input{recovering_product_analysis}
\subsection{Smooth integers in short intervals}
\label{secsec:prelim}

We review a few facts about smooth integers 
and refer to Andrew Granville's survey~\cite{Granville_smoothnumbers:}
for details and proofs.
Let $S(x,y)$ be the set of integers up to $x$, all of whose prime factors 
are less or equal to $y$
(such integers are called ``$y$-smooth''), 
and let ${\Psi}(x,y)$  be the number of such integers.
It is a remarkable result that for any fixed $u \geq  1$, the
proportion of the integers up to $x$, that only have prime 
factors less or equal to $x^{1/u}$,  tends to a
nonzero limit as $x$ escapes to infinity.
This limit, denoted by ${\rho}(u)$ is known as the 
Dickman-de Bruijn ${\rho}$-function.
To be precise, Dickman's result states the following
asymptotic equivalence
\begin{equation}
{\Psi}(x,y) \ \sim  \  x \, {\rho}(u) \ \ {\rm as} \ \ x \longrightarrow \infty
\end{equation}
where $x = y^u$.
Unfortunately, one cannot write down a useful, 
simple function that gives the value of ${\rho}(u)$
for all $u$.
\iflongversion
Therefore, upper bounds and lower bounds
for ${\Psi}(x,y)$ are of interest, in particular
for the purpose of analyzing algorithms where smooth integers play a
key role, as for the multiplication algorithm of
Section~\ref{sec:cmmxxmultiplication}.
A simple lower bound is given by 
\begin{equation}
{\Psi}(x,y) \ \geq \ x^{1 - {\log}({\log}(x)) / {\log}(y)}
\end{equation}
for all $x \geq y \geq 2$ and $x \geq 4$,
from which, one immediately deduces
\begin{equation}
{\Psi}(x,{\log}^2(x)) \ \geq \ \sqrt{x}.
\end{equation}
\fi 
However, it is desirable to obtain statements 
which could imply inequalities like 
${\Psi}(x+z,y) - {\Psi}(x,y) > 0$
for all $x \geq z$ and for $y$ arbitrary small.
Indeed, such inequality would mean that, 
for all $x,y$, a ``short interval'' 
around $x$ contains at least one $y$-smooth integer.
\iflongversion
In fact, the following relation is well-known for all $x \geq z \geq x/y^{1 - o(1)}$:
\begin{equation}
{\Psi}(x+z,y) - {\Psi}(x,y) \ \sim \ \frac{z}{x} {\Psi}(x,y) \ \sim \ 
                            z {\rho}(u).
\end{equation}
but does not meet our needs
since it does allow us to make  $y$ arbitrary small
independently of $x$ and $z$,
while implying ${\Psi}(x+z,y) - {\Psi}(x,y) > 0$.

Nevertheless, in 1999, Ti Zuo Xuan~\cite{1999-xuan}
\else
In 1999, Ti Zuo Xuan~\cite{1999-xuan}
\fi
proved that, under the Riemann Hypothesis (RH), 
for any ${\varepsilon} > 0, \ {\delta} > 0$ 
there exists $x_0$ such that for all $x \geq x_0$
the interval $(x, x + y]$,
for $\sqrt{x} ({\log}(x))^{1 + \epsilon} \leq y \leq x$,
contains an integer having no prime factors exceeding $x^{\delta}$.
Moreover, in 2000, Granville has conjectured 
that for all ${\alpha} > 0$ there exists $x_0$ such that for all $x \geq x_0$,
we have
\begin{equation}
\label{eq:Granville}
{\Psi}(x + \sqrt{x},x^{\alpha}) - {\Psi}(x,x^{\alpha}) > 0.
\end{equation}
In the sequel of this section, we shall assume that 
either RH or Granville's conjecture holds.
In fact, we have verified Relation~(\ref{eq:Granville})
experimentally for all $x \leq 2^{23}$.
This is by far sufficient for the purpose of 
implementing our multiplication algorithm
and verifying the properties of the positive integers $N, K, M$
stated in Theorem~\ref{thrm:complexity}.
Moreover, for all the values $x$ that we have tested,
the following holds
\begin{equation}
\label{eq:GranvilleTwo}
{\Psi}(2x, 23) - {\Psi}(x, 23) > 0.
\end{equation}

\fi

\ifnewversion
\subsection{Proof of Theorem~\ref{thrm:complexity}}
\label{sec:proofoftheoremone}
Recall that analyzing our algorithm reduces to 
analyzing Lines {\bf 2}, {\bf 5}, {\bf 7}.
Based on the results obtained above
for $W_2$, $W_5$, $W_7$, 
with Relations (\ref{eq:WTwo}), (\ref{eq:WFive}), (\ref{eq:WSeven}), 
respectively, it follows
that the estimate for the work of the whole algorithm
is given by $W_5$, leading to the result in Theorem~\ref{thrm:complexity}.
Meanwhile, the span of the whole algorithm is dominated by $S_7$ and
one obtains the result in Theorem~\ref{thrm:complexity}.
Finally, the cache complexity estimate
in Theorem~\ref{thrm:complexity} comes from adding up 
$Q_2$, $Q_5$, $Q_7$ and simplifying.

\else
\input{proof_of_theorem}
\fi

\fi

\section{Experimentation}
\label{sec:experimentation}

We realized an implementation of a modified version
of the algorithm presented in Section~\ref{sec:cmmxxmultiplication}.
The only modification is that we rely on prime numbers $p$
that do not satisfy Equation (\ref{eq:Fes}).
\correction{1}{In particular, our implementation is not using yet 
generalized Fermat primes. Overcoming this limitation is work in progress.
Currently, our prime numbers are of machine word size.
As a consequence, the work of the implemented algorithm
is $O(d K M {\log}(d K) ({\log}(d K) + 2 M))$,
which is asymptotically larger than the work of the approach
combining Kronecker's substitution and Sch{\"o}nhage \& Strassen.
This helps understanding why the speedup factor 
of our parallel implementation over the integer polynomial
multiplication code of FLINT (which is a serial implementation of
the algorithm of Sch{\"o}nhage \& Strassen)
is less than the number of cores that we use.}

However, this latter complexity estimate yields
a (modest) optimization trick: since the asymptotic
upper bound $O(d K M {\log}(d K) ({\log}(d K) + 2 M))$
increases slower with $M$ than with $d$ or $K$,
it is advantageous to make $M$ large. We do that
by using two machine word primes (and the Chinese
Remaindering Algorithm) instead of using a single prime
for computing two modular images of $C^{+}({x,y})$ and $C^{-}({x,y})$ 
that we combine by the Chinese Remaindering Algorithm.
Moreover, our parallel implementation outperforms the integer polynomial
multiplication code of {\sc Maple} 2015. 

Our code is written in the multi-threaded
language {\tt CilkPlus}~\cite{DBLP:journals/tjs/Leiserson10} and compiled 
with the {\tt CilkPlus} branch of
GCC. 
Our experimental results were obtained on a 48-core AMD Opteron 6168,
running at 900MHz with 256 GB of RAM and 512KB of L2 cache.
Table~\ref{classical4andCVL2} gives running times for the five  multiplication
algorithms that we have implemented:
\begin{itemizeshort}
\item  {\small ${\rm KS}_s$} stands for
Kronecker's substitution combined with Sch{\"o}nhage \& Strassen algorithm~\cite{DBLP:journals/computing/SchonhageS71};
note this is a serial implementation, running on  1 core.
\item ${\rm CVL}_p^2$ is the prototype implementation of 
the modified version of the algorithm described in 
Section~\ref{sec:cmmxxmultiplication}, running on 48 cores.
In this implementation, two machine-word size Fourier primes are used
instead of a single big prime for computing $C^{+}({x,y})$ and $C^{-}({x,y})$.
\item {\small ${\rm DnC}_p$} is a straightforward 4-way divide-and-conquer parallel implementation of plain 
multiplication, run on 48 cores, see Chapter 2 of~\cite{farnamThesis}.
\item {\small ${\rm Toom}^4_p$} is a parallel implementation of 4-way Toom-Cook, run on 48 cores, see Chapter 2 of~\cite{farnamThesis}.
\item {\small ${\rm Toom}^8_p$}  is a parallel implementation of 8-way Toom-Cook, run on 48 cores, see Chapter 2 of~\cite{farnamThesis}. 
\end{itemizeshort}
In  Table~\ref{classical4andCVL2},
for each example, 
the degree $d$ of the input polynomial is equal to the coefficient bit size $N$.
The input polynomials $a(y), b(y)$  are random and dense.

\begin{figure}[htb]
\centering
\includegraphics[width=.9\linewidth]{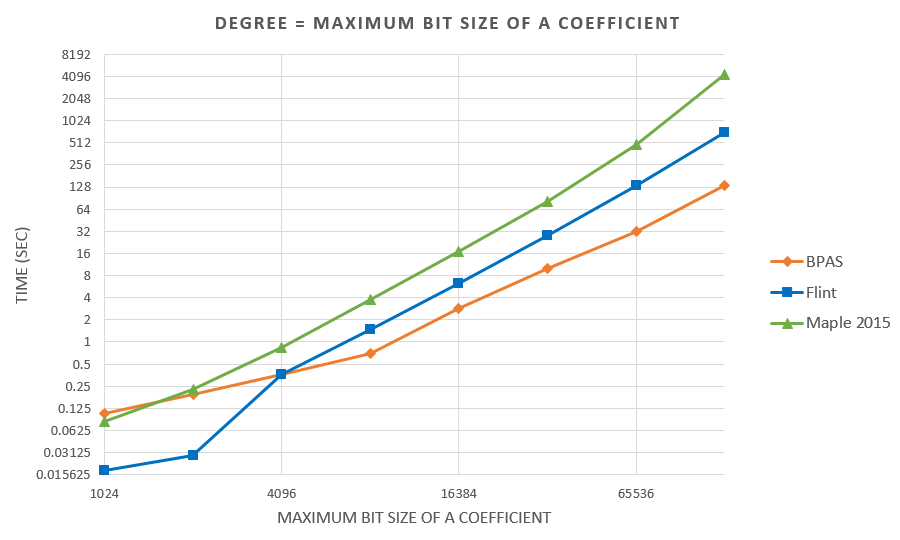}

\includegraphics[width=.9\linewidth]{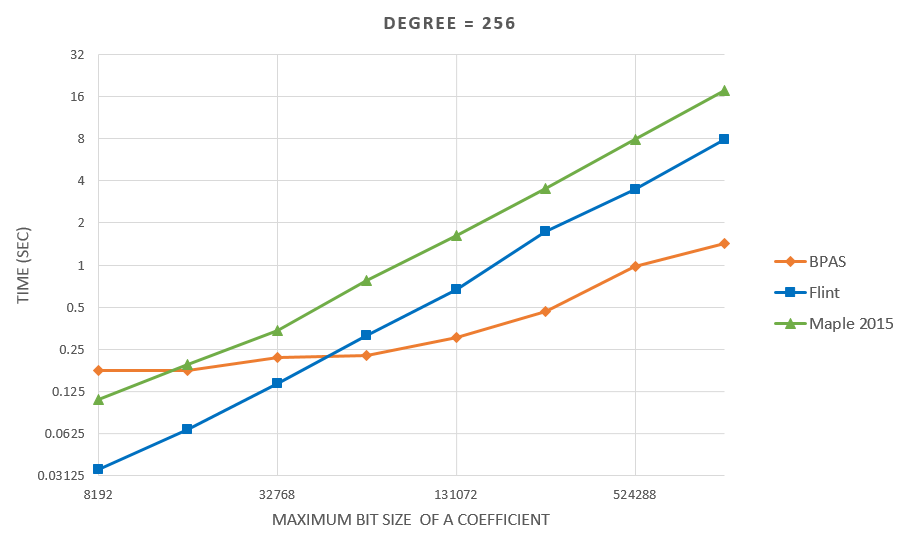}

\caption{\small Dense integer polynomial multiplication: ${\rm CVL}_p^2$ (BPAS) vs FLINT vs {\sc Maple} 2015}
\label{fig:BPASbench}
\end{figure}

Figure~\ref{fig:BPASbench} shows the running time comparison
among our algorithm, FLINT 2.4.3~\cite{Hart11flintfast} and {\sc Maple} 2015.
The input of each
test case is a pair of polynomials of degree $d$ where each
coefficient has bit size $N$. 
Timings (in sec.) appear along the vertical axis. 
Two plots are provided: one for which $d = N$ holds and one for $d$ is
much smaller than $N$.

From Table~\ref{classical4andCVL2} and Figure~\ref{fig:BPASbench}, 
we observe that our implementation ${\rm CVL}_p^2$ performs
better on sufficiently large
input data, compared to its counterparts.

 \ifcomment
\else

\begin{table}[htbp]
\centering
\ifthesis
\else
\scriptsize
\fi
\begin{tabular}{|c|c|c|c|c|c|c|}
\hline
  {\tiny $d, N$} & {\tiny ${\rm CVL}_p^2$} &{\tiny ${\rm DnC}_p$} &{\tiny ${\rm Toom}^4_p$} &{\tiny ${\rm Toom}^8_p$} & {\tiny ${\rm KS}_s$}  \\
	\hline
	$2^9$     & 0.152 & 0.049 &0.022  &0.026 & 0.018 \\
	$2^{10}$    & 0.139 & 0.11  &0.046  &0.059 & 0.057 \\
	$2^{11}$    & 0.196 & 0.17  &0.17   &0.17 & 0.25 \\
	$2^{12}$    & 0.295 & 0.58  &0.67   &0.64 & 1.37 \\
	$2^{13}$    & 0.699 & 2.20  &2.79   &2.73 & 5.40 \\
	$2^{14}$   & 1.927 & 8.26  &10.29  &8.74  & 20.95 \\
	$2^{15}$   & 9.138 &30.75  &35.79  &33.40 & 92.03 \\
	$2^{16}$   & 33.04 &122.1 &129.4 &115.9 & *Err.  \\
	\hline
\end{tabular}
\caption{\small Timings of polynomial multiplication with $d=N$}
\label{classical4andCVL2}
\end{table}

\fi

\section{Concluding Remarks}
\label{sec:discussion}

We have presented a parallel FFT-based method for
multiplying dense univariate polynomials with integer
coefficients. 
Our approach relies on two convolutions (cyclic and negacyclic)  of
bivariate polynomials which allow us to take advantage
of the row-column algorithm for 2D FFTs.
The proposed algorithm is asymptotically faster than that
of Sch{\"o}nhage \& Strassen.

In our implementation, 
the data conversions between univariate polynomials over {\Z}
and bivariate polynomials over ${\Z}/p{\Z}$  are highly optimized
by means of  low-level ``bit hacks''  thus avoiding software
multiplication of large integers.
In fact, our code relies only and directly on machine word operations
(addition, shift and multiplication).

Our experimental results show this new algorithm has a high parallelism
and scales better than its competitor algorithms.
Nevertheless, there is still room for improvement
in the implementation. Using a single big prime (instead
of two machine-word size primes and the Chinese Remaindering Algorithm),
and requiring that this prime be a generalized Fermat prime,
would make the implemented algorithm follow strictly
that presented in Section~\ref{sec:cmmxxmultiplication}.

\nocite{DBLP:conf/issac/GaudryKZ07} 
\nocite{DBLP:conf/csr/Pospelov11}
\nocite{DBLP:journals/computing/SchonhageS71}
\nocite{DBLP:journals/siamcomp/Furer09} 
\nocite{DBLP:journals/siamcomp/DeKSS13} 
\nocite{DBLP:conf/stoc/DeKSS08}
\nocite{DBLP:conf/casc/MengJ13}
\nocite{DBLP:conf/cap/MengVJMFX10} 
\nocite{Harvey:2009:FPM:1568787.1568920} 
\nocite{DBLP:conf/issac/BodratoZ07} 

\section*{Acknowledgments}
The authors would like to thank IBM Canada Ltd and
NSERC of Canada for supporting their work.

\bibliographystyle{abbrv}
\bibliography{reference}

\end{document}